\documentclass[prb,preprint,superscriptaddress,groupedaddress,showpacs,preprintnumbers,amsmath,amssymb]{revtex4}
\usepackage{graphicx}% Include figure files
\usepackage{dcolumn}% Align table columns on decimal point
\usepackage{bm}% bold math

\begin{document}

%\preprint{}
%%\preprint{Phys. Rev. B}

\title{Comments on ``Spin-dependent tunneling through a symmetric semiconductor 
barrier: The Dresselhaus effect''}


\author{Titus Sandu}
 \email{titus.sandu@umontreal.ca}
\affiliation{D\'{e}partement de chimie, Universit\'{e} de Montr\'{e}al, 
C.P. 6128, succursale Centre-ville, Montr\'{e}al, Qu\'{e}bec H3C 3J7, Canada
 }%
%\}%
%\author{}
% \email{}
%  \homepage{}
%\affiliation{%
%\\
%  }%
  
%\author{}%
% \email{}
%\affiliation{
%\\
%\\}%

%\author{}%
% \email{}
%\affiliation{}%
%\author{}
%\thanks{}
%\affiliation{}


\date{\today}
\begin{abstract}
In a recent paper [Phys. Rev. B 72, 153314 (2005)], the $k^{3}$-Dresselhaus 
term in the contacts and the full form of the current operator are 
considered for spin-dependent tunneling 
through a symmetric barrier. The authors found that the full form of the current operator 
has a much larger influence on the spin polarization than it was initially 
thought. 
In this Comment we will show that their treatment of the other problem, the $k^{3}$-Dresselhaus 
term in the contacts, is 
incorrect. Their proposed solution in the contacts simply does not obey the Schr\"{o}dinger equation.
In this context we also comment on the definition and the suitability of spin polarization in contacts 
with spin-orbit coupling. 
\end{abstract}

%\keywords{}

\maketitle

%\section{}
%\subsection{}
%\subsubsection{}

%\section{Introduction}
Spin-dependent tunneling through a symmetric barrier has made the object of 
recent investigations.\cite{Perel03, Wang05} The first paper\cite{Perel03} 
considered the Dresselhaus term only 
in the barrier and neglected the small corrections induced to the effective masses in the current 
operator. The second paper\cite{Wang05} added the $k^{3}$-Dresselhaus term 
in the contact regions and considered the full expression of the current 
operators. The authors of the second paper\cite{Wang05} show that the full form of the 
current operator changes substantially the spin polarization. However, their treatment of the
the $k^{3}$-Dresselhaus term is not satisfactory: the solution proposed by the authors\cite{Wang05} 
does not obey the Schr\"{o}dinger equation in the contacts. Below we will show that. 

Beside the formal aspect of the solution presented in Ref. \onlinecite{Wang05}, 
there is also a practical aspect. 
As the authors have found, the small spin-dependent corrections to the current 
operator will induce large variations in the spin polarization (Figs. 2 and 3 in their paper).\cite{Wang05} 
These additional terms, which 
are around $2\%$ corrections to the effective masses for the spin-dependent 
tunneling, translate into the variation of spin polarization by more 
than $30\%$  (Fig. 2 in Ref. \onlinecite{Wang05}). These huge variations 
in polarization are due 
to the fact that the entire mechanism is driven by tunneling which has 
an exponential dependence on parameters.\cite{Perel03} 
In addition, the authors' solution in Ref. \onlinecite{Wang05} will be quite different than 
the full solution presented below due to new terms that are missing from the boundary 
conditions at the barrier. Therefore, if the solution presented in Ref. \onlinecite{Wang05} 
was to be considered as an approximation, one would expect not only quantitatively but also qualitatively large 
variations with respect to the full calculation 
(i.e. with the true solution of the Schr\"{o}dinger equation as it will be outlined below). 

We consider the transmission of an electron with the 
wave vector $k = \left( {k_{\vert \vert } ,k_z } \right)$ through a barrier 
$V\left( z \right)$ along z-axis.
Below, we, 
basically, use the same notations. The electron Hamiltonian in the 
effective mass approximation is

\begin{equation}
\label{eq:H0}
H_0 = \frac{P^2}{2m^\ast } + V\left( z \right) + H_D. 
\end{equation}
\noindent
$H_D$ is the spin-dependent Dresselhaus Hamiltonian with the expressions in the 
contact regions as

\begin{equation}
\label{eq:DV}
H_D = \gamma _1 \left[ {k_x \left( {k_y^2 - k_z^2 } \right)\sigma _x + k_y 
\left( {k_z^2 - k_x^2 } \right)\sigma _y + k_z \left( {k_x^2 - k_y^2 } 
\right)\sigma _z } \right]
\end{equation}

\noindent
and in the barrier as

\begin{equation}
\label{eq:DB}
H_D = \gamma _2 \left( {k_x \sigma _x - k_y \sigma _y } 
\right)\frac{\partial ^2}{\partial ^2z^2}.
\end{equation}

The two-component solution is found by solving separately the 
Schr\"{o}dinger equation in the contact and barrier regions and matching the 
wave function and the current at the interfaces. In the left contact the 
solution should have the form of an incoming wave in addition to a scattered (reflected) 
wave \cite{Belkic} in the spinor states in which the motion of the electron 
becomes free-like. 
This is achieved by the spinor eigenvectors of the Hamiltonian~(\ref{eq:DV}). 
In the right contact, the solution should be an outgoing wave in the same 
spinor eigenstates of the Hamiltonian~(\ref{eq:DV}). In the barrier, the solution is 
cast in the spinor eigenstates of the Hamiltonian~(\ref{eq:DB}). 
We denote by $\left| {\eta _\pm } \right\rangle $, the spinor eigenvectors 
of the Hamiltonian (\ref{eq:DV}) in the contacts and by $\left| {\chi _\pm } \right\rangle 
$, the spinor eigenvectors of the Hamiltonian (\ref{eq:DB}) in the barrier. 
Thus the formal solution is

\begin{equation}
\label{eq:solution}
\begin{array}{l}
 \left| {\Psi _L } \right\rangle = \exp \left( {ik_{\vert \vert } \cdot \rho 
} \right)\sum\limits_{j = \pm } {\left[ {\exp \left( {ik_j z} \right) + r_j 
\exp \left( { - ik_j z} \right)} \right]\,\left| {\eta _j } \right\rangle } 
, \\ 
 \left| {\Psi _B } \right\rangle = \exp \left( {ik_{\vert \vert } \cdot \rho 
} \right)\sum\limits_{j = \pm } {\left[ {A_j \exp \left( {q_j z} \right) + 
B_j \exp \left( { - q_j z} \right)} \right]\,\left| {\chi _j } 
\right\rangle } , \\ 
 \left| {\Psi _R } \right\rangle = \exp \left( {ik_{\vert \vert } \cdot \rho 
} \right)\sum\limits_{j = \pm } {\left[ {t_j \exp \left( {ik_j z} \right)} 
\right]\,\left| {\eta _j } \right\rangle } . \\ 
 \end{array}
\end{equation}

\noindent
In Eq.~(\ref{eq:solution}), $k_\pm \left( {q_\pm } \right)$ are the wave vectors associated with the 
solution of the Schr\"{o}dinger equation outside (inside) the barrier along the 
eigenstate spinors $\left| {\eta _\pm } \right\rangle \left( {\left| {\chi 
_\pm } \right\rangle } \right)$. Equation (\ref{eq:solution}) is expressed for the left contact, 
the barrier, and the right contact, respectively. We would like to mention an analogous model of the atom-atom scattering 
where the atoms have two internal states.\cite{Kokkelmans02} 

In Ref.~\onlinecite{Wang05}, however, the 
authors cast the solution in the spinor basis $\left| {\chi _\pm } \right\rangle$ 
across the whole structure. Their solution reads 

\begin{equation}
\label{eq:solWang}
\begin{array}{l}
 \left| {\Psi '_L } \right\rangle = \exp \left( {ik_{\vert \vert } \cdot 
\rho } \right)\sum\limits_{j = \pm } {\left[ {\exp \left( {ik_j z} \right) + 
r'_j \exp \left( { - ik_j z} \right)} \right]\,\left| {\chi _j } 
\right\rangle } , \\ 
 \left| {\Psi '_B } \right\rangle = \exp \left( {ik_{\vert \vert } \cdot 
\rho } \right)\sum\limits_{j = \pm } {\left[ {A'_j \exp \left( {q_j z} 
\right) + B'_j \exp \left( { - q_j z} \right)} \right]\,\left| {\chi _j } 
\right\rangle } , \\ 
 \left| {\Psi '_R } \right\rangle = \exp \left( {ik_{\vert \vert } \cdot 
\rho } \right)\sum\limits_{j = \pm } {\left[ {t'_j \exp \left( {ik_j z} 
\right)} \right]\,\left| {\chi _j } \right\rangle } . \\ 
 \end{array}
\end{equation}

\noindent
This solution [Eq.~(\ref{eq:solWang})] given by Wang {\it et al.}~\cite{Wang05} does 
not obey the Schr\"{o}dinger equation in the 
contacts for a scattering/transport problem. 
To show that, we apply the spinor identity operator 
written as $ \sum\limits_{j = \pm } {\left| {\eta _j } 
\right\rangle \left\langle {\eta _j } \right|} $ on the solution ~(\ref{eq:solWang}) 
in the contact regions. It is easy to notice that, for example, the component of $\left| {\Psi '_L } \right\rangle 
$ that is proportional to $\left| {\eta _ + } \right\rangle $, 

\begin{widetext}
\begin{equation}
\label{eq:componentwf}
\left\langle {\eta _ + } \mathrel{\left| {\vphantom {{\eta _ + } {\chi _ + 
}}} \right. \kern-\nulldelimiterspace} {\chi _ + } \right\rangle e^{ik_{ + 
\,} z} + \left\langle {\eta _ + } \mathrel{\left| {\vphantom {{\eta _ + } 
{\chi _ - }}} \right. \kern-\nulldelimiterspace} {\chi _ - } \right\rangle 
e^{ik_{ - \,} z} + r'_ + \left\langle {\eta _ + } \mathrel{\left| {\vphantom 
{{\eta _ + } {\chi _ + }}} \right. \kern-\nulldelimiterspace} {\chi _ + } 
\right\rangle e^{ - ik_{ + \,} z} + r'_ - \left\langle {\eta _ + } 
\mathrel{\left| {\vphantom {{\eta _ + } {\chi _ - }}} \right. 
\kern-\nulldelimiterspace} {\chi _ - } \right\rangle e^{ - ik_ - \,z},
\end{equation}
\end{widetext}

\noindent
does not satisfy the Schr\"{o}dinger equation in the spin channel $\left| {\eta _ + } 
\right\rangle $ due to the terms that contain the factors $e^{ \pm ik_{ 
- \,} z}$. The factors $e^{ \pm ik_{ - \,} z}$ are associated with the Schr\"{o}dinger equation in the 
spin channel $\left| 
{\eta _ - } \right\rangle $. Similar analysis can be done for the outgoing channels. 
Thus the wave vector (\ref{eq:solWang})  used
 in Ref.~\onlinecite{Wang05} is not a solution for the spin-dependent transport problem 
across the barrier with the $k^{3}$-Dresselhaus term in the contacts. Their solution 
mixes the spin states ``+'' with the spin states ``-'' in the incoming and outgoing channels. 

Finally, we comment on the polarization. Spin polarization is calculated with the assumption that 
a non-polarized spin current is injected into the system. In Ref.~\onlinecite{Wang05}, the spin polarization 
of the barrier was calculated along the spin direction given by the 
spinors $\left| {\chi _\pm } \right\rangle $. However, the spin polarization along 
$\left| {\chi _\pm } \right\rangle $ derived from solution ~(\ref{eq:solution}) can 
be easily obtained as 

\begin{widetext}
\begin{equation}
\label{eq:component}
P' =  \\
 \frac{\left| {t_ + } \right|^2\left( {\left| {\left\langle {\chi _ + } 
\mathrel{\left| {\vphantom {{\chi _ + } {\eta _ + }}} \right. 
\kern-\nulldelimiterspace} {\eta _ + } \right\rangle } \right|^2 - \left| 
{\left\langle {\chi _ - } \mathrel{\left| {\vphantom {{\chi _ - } {\eta _ + 
}}} \right. \kern-\nulldelimiterspace} {\eta _ + } \right\rangle } 
\right|^2} \right) + \left| {t_ - } \right|^2\left( {\left| {\left\langle 
{\chi _ + } \mathrel{\left| {\vphantom {{\chi _ + } {\eta _ - }}} \right. 
\kern-\nulldelimiterspace} {\eta _ - } \right\rangle } \right|^2 - \left| 
{\left\langle {\chi _ - } \mathrel{\left| {\vphantom {{\chi _ - } {\eta _ - 
}}} \right. \kern-\nulldelimiterspace} {\eta _ - } \right\rangle } 
\right|^2} \right) + \left( 
{\mbox{z}\;\mbox{dep.}\;\mbox{terms}} 
\right)}{\left| {t_ + } \right|^2 + \left| {t_ - } \right|^2}
\end{equation}
\end{widetext}

\noindent
Equation~(\ref{eq:component}) tells us that, due to mixing, the polarization measured along the 
spinor $\left| {\chi _\pm } \right\rangle$ is different from the one measured along 
$\left| {\eta _\pm } \right\rangle $. It is also dependent on the $z$ coordinate with a close 
resemblance to optical activity or/and birefringence, where the rotation of the light 
polarization vector is related to the length of the light path. Therefore, 
the polarization can be well defined in a unique way in the $\left| {\eta _\pm } 
\right\rangle $-spinor representation. Nevertheless, the spin polarization is well defined in the 
spinor basis $\left| {\chi _\pm } \right\rangle$ only if the $k^{3}$-Dresselhaus 
term in the contacts is neglected or zero.
We conclude that the solution to the spin transport through a symmetric barrier with $k^{3}$-Dresselhaus 
term in the contacts as it was presented in Ref.~\onlinecite{Wang05} is incorrect due to inappropriate resolution 
of the Schr\"{o}dinger equation in the contacts.

\end{document}